\documentclass[reprint,
superscriptaddress,
amsmath,amssymb,
aps,prd,twocolumn,longbibliography,
floatfix,
]{revtex4-1}

\usepackage{lineno}
\usepackage{mathtools}
\usepackage{graphicx}% Include figure files
\usepackage{dcolumn}% Align table columns on decimal point
\usepackage{bm}% bold math
\usepackage{float}
\usepackage{xcolor}
\usepackage{algorithm}
\usepackage{algpseudocode} % in place of algorithmic. Conflicts with revtex4 unless [H] option passed
\usepackage{xspace}
\usepackage{url}
\usepackage[colorlinks=true, citecolor=cyan, linkcolor=cyan, urlcolor= cyan]{hyperref} 
\newcommand{\beq}{\begin{equation}}
\newcommand{\eeq}{\end{equation}}
\newcommand{\bea}{\begin{eqnarray}}
\newcommand{\eea}{\end{eqnarray}}
\newcommand{\ba}{\begin{array}}
\newcommand{\ea}{\end{array}}
\newcommand{\sqz}{\medmuskip=0mu}

\newcommand{\mlam}{{\bm{\lambda}}}
\newcommand{\mLam}{{\bm{\Lambda}}}
\newcommand{\bilby}{{\sc Bilby}\xspace}
\newcommand{\lal}{{\sc LAL}\xspace}
\newcommand{\lalinference}{{\sc LALInference}\xspace}
\newcommand{\lalsuite}{{\sc LALSuite}\xspace}
\newcommand{\lalsimulation}{{\sc LALSimulation}\xspace}
\newcommand{\python}{{\sc Python}\xspace}
\newcommand{\pyroq}{{\sc PyROQ}\xspace}
\newcommand{\greedycpp}{{\sc GreedyCpp}\xspace}
\newcommand{\gstlal}{{\sc GstLAL}\xspace}
\newcommand{\pycbc}{{\sc PyCBC}\xspace}
\newcommand{\pip}{{\sc PIP}\xspace}

\begin{document}

\title{PyROQ: a Python-based Reduced Order Quadrature Building Code for Fast Gravitational Wave Inference}

\author{Hong Qi}
\email{hong.qi@ligo.org}
\author{Vivien Raymond}
\affiliation{School of Physics and Astronomy, Cardiff University, Cardiff, UK, CF24\ 3AA }
%\collaboration{LIGO-Virgo Collaboration}

\date{\today}% It is always \today, today,

\begin{abstract}

The next generation of gravitational-wave observatories will reach low frequency limits on the orders of a few Hz, thus enabling the detection of gravitational wave signals of very long duration. The run time of standard parameter estimation techniques with these long waveforms can be months or even years, making it impractical with existing Bayesian inference pipelines. Reduced order modeling and reduced order quadrature integration rule have recently been exploited as promising techniques that can greatly reduce parameter estimation computational costs. We describe a \python-based reduced order quadrature building code, \pyroq, which builds the reduced order quadrature data needed to accelerate parameter estimation of gravitational waves. We present the first bases for the IMRPhenomXPHM waveform model of binary-black-hole coalescences, including subdominant harmonic modes and precessing spins effects. Furthermore, the code infrastructure makes it directly applicable to the gravitational wave inference for space-borne detectors such as the Laser Interferometer Space Antenna (LISA). %In addition, the techniques are broadly applicable to other research fields where fast Bayesian analysis is required.
%\begin{description}
%\item[Keywords]
%ROQ, PyROQ, Gravitational Waves, Parameter Estimation
%\end{description}
\end{abstract}

\maketitle

%\tableofcontents
\section{Introduction}

Ground-based gravitational-wave observatories such as LIGO \citep{aasi2015}, Virgo \citep{acernese2015}, and KAGRA \citep{kagra2019} are now frequently detecting gravitational wave (GW) signals from the mergers of binary black holes (BBHs) \citep{gwtc-1, gwtc-2, GW190412, GW190425, GW190521, first2nsbh}, binary neutron stars (BNSs) \citep{GW190814} and neutron star - black hole binaries (NSBHs) \citep{GW190814, S200105ae-gcn, S200115j-gcn}. With the ever increasing sensitivity of new generation detectors expected to come online in the near future (e.g., A+ \citep{LIGOScientific:2019vkc}, Voyager \citep{Voyager1, Voyager2, Voyager3}, the Einstein Telescope \citep{ET2010}, and Cosmic Explorer \citep{CE2019}), we will start detecting gravitational waves at lower frequencies \citep{lowfreq13,Yu:2017zgi}, with the low frequency limit going from the current 10 Hz down to 5 Hz or even 1Hz. In order to take full advantage of this sensitivity, waveform models are required to span the whole frequency range, starting from the lowest frequency possible. It can take a unfeasible length of time to compute the likelihood for such signals, unless efficient representation techniques are used, such as Singular Value Decomposition \citep{kipp2010,smith2013,Constantine2014}, Reduced Order Modeling (ROM) \citep{Field2013, Canizares2014, Smith2016, Field2018}, Relative Binning \citep{Zackay:2018qdy}, Multi-banding \citep{Vinciguerra:2017ngf} and likelihood-free approaches \citep{Gabbard:2019rde, Chua:2019wwt, Green:2020hst}. %\hq{\textit{reduced order quadrature} (ROQ) rule}
In this work we focus on the application of ROM technique as applied to the calculation of the likelihood integrand, building the \textit{reduced order quadrature} (ROQ) rule. For the longest waveform signal used, speedup of several orders of magnitude is possible. The main computational cost is moved to an offline basis building stage, which can be done once per waveform model and ahead of time. However, as new and more complex waveform models are being developed at a faster rate \citep{PhenomNSBH,IMRPhenomXHM,IMRPhenomTP,IMRPhenomXPHM}, a user-friendly and easy-to-use ROQ basis building code developed in \python (the language of choice of the LIGO-Virgo-KAGRA (LVK) community and wider gravitational-wave community) is desirable in order to fully exploit the more accurate models.

In addition, the most advanced waveform models that enable the extraction of the most astrophysical information from gravitational-wave signals are also the most computationally expensive. Many of the most impactful studies will involve waveform models of more detailed physics, such as precession and subdominant harmonic mode effects. The high computational costs associated with performing parameter estimation with these longer, more physically accurate waveform models, make it almost impossible with standard Bayesian inference pipelines. For example, it can take several months to finish a parameter estimation run on a BNS signal using standard methods, whereas techniques like ROQ rule can reduce the run time to about 24 hours \citep{Smith2016, Dai:bnspe} and under certain conditions  a couple of hours \citep{soichiro}. Furthermore, in practice re-analyses are often needed due to fine-tuning and errors, as well as using different stages and types of calibrated data such as cleaned data \citep{gwtc-1}. %Active efforts have been made on reduced order modeling (ROM) and reduced order quadrature (ROQ) rules for parameter estimation of gravitational waves \cite{}. 

In this paper, we present a novel method to search for reduced bases and construct ROQ data for gravitational waveforms. We also showcase the implementation of the method, i.e., the \pyroq code written in \python. Our effort is complimentary to the ground-breaking C++ code \greedycpp \citep{Field2013, Field2018} for offline ROQ constructions, which uses a greedy algorithm to search for waveform bases. We show that even though \python is in general slower than C++ (see for instance \citep{hope2015}), the novel algorithm implemented in \pyroq enables the waveform bases to be produced in comparable times to the efficient \greedycpp code. We also illustrate that \pyroq can build bases of smaller sizes at a better interpolation accuracy compared to \greedycpp.

The paper is organized as follows. In Sec.~\ref{sec:methodology} we introduce the basics of the ROQ rule for gravitational wave parameter estimation, the requirements on reduced bases, and the algorithm \pyroq uses to search for reduced bases and produce ROQ data. In Sec.~\ref{sec:codevalidation} we describe the validation of the \pyroq code by applying it to the {\tt IMRPhenomPv2} \citep{IMRPhenomPv2-1, IMRPhenomPv2-2} waveform model and checking against two sets of ROQs built with \greedycpp. The interpolation accuracy of \pyroq built bases is also thoroughly illustrated. In Sec.~\ref{sec:applications} we first use \bilby pipeline \citep{Ashton:2018jfp} to show the likelihood comparisons and the posterior comparisons, respectively, for the recoveries of simulated NSBH injections into zero noise with the standard method and the ROQ method, where the ROQ data were built with \pyroq. We then compare the inferences on the first detected BBH merger GW150914 using the two methods with \pyroq built ROQs. We demonstrate new applications of \pyroq with its first constructed ROQs for one of the most advanced BBH waveform model {\tt IMRPhenomXPHM} for LIGO-Virgo-KAGRA detectors and supermassive black hole binaries (SMBHBs) in the context of Laser Interferometer Space Antenna (LISA). %We also briefly introduce the application of \pyroq in the LISA gravitational wave inference with supermassive black hole binaries (SMBHBs). 
Finally, we conclude in Sec.~\ref{sec:conclusion}. 

The \pyroq code is publicly available at \url{http://pypi.org/project/PyROQ} although there can be some delays between the development and the releases. It can easily be installed with \pip.

%%%%%%%%%%%%%%%%%%%%%%%
\section{Methodology}
\label{sec:methodology}
\subsection{Basics of inference on gravitational wave}
\label{subsec:peintroduction}
Gravitational-wave inference provides the  probability density function (PDF) of a set of parameters $\mLam$ parameterizing a model of the gravitational-wave signal, $h(\mLam)$, given the detector's collected strain data $d$. According to Bayes’ theorem, the PDF is given by

\beq
p(\mLam| d)  = \frac{\mathcal{P}(\mLam)\ \mathcal{L}(d | \mLam)}{\mathcal{E}(d)}\,, \label{eq:posterior}
\eeq
where $p(\mLam | d)$ is the posterior probability of the model parameters given the data, $\mathcal{P}( \mLam )$ is the prior probability on the model parameters, $\mathcal{L}(d | \mLam)$ is the likelihood of the data at given model parameters, and $\mathcal{E}(d)$ is the model evidence which describes the probability of the data given the model. For a certain hypothesis, the evidence does not depend on the parameters $\mLam$, and thus enters only as an overall scaling factor in parameter estimation. The prior is a parameterization of the a priori knowledge of the parameters, and can often be written analytically. The likelihood is in general the most computationally costly quantity.

Suppose the detector data $d$ is composed of a gravitational wave signal $h(\mLam_{\tt true})\equiv h(t;\mLam_{\tt true})$ and noise $n$, i.e., $ d= h(\mLam_{\tt true}) +  n$. The log-likelihood function can be computed by
\begin{align}
\log\mathcal{L}(d|\mLam) &= - \frac{1}{2} (d - h(\mLam), d - h(\mLam)) \, ,  \label{eq:likelihood} 
\end{align}
where  $(a, b)$ is an \textit{overlap} integral approximated in a discrete form
\beq
(d, h(\mLam)) = 
4\Re\ \Delta f \sum_{k=1}^{L} \frac{\tilde d^*(f_k)\tilde h(f_k;\mLam)}{S_{n}(f_k)}\, . \label{eq:inner}
\eeq
The $\tilde d(f_k)$ and $\tilde h(f_k;\mLam)$ are the discrete Fourier transforms of $d(t)$ and $h(t;\mLam)$ at frequencies $\{ f_k \}_{k=1}^L$ and $S_{n}(f_k)$ is the power spectral density (PSD) of the detector's noise. For a given observation time $T=1/\Delta f$ and detection frequency window from $f_{\tt low}$ to $f_{\tt high}$, there are $L \sim {\tt int}\left( \left[ f_{\tt high} - f_{\tt low} \right] T \right)$ sampling points in Eq.~(\ref{eq:inner}).
When $L$ is large, the evaluation of the model at each $f_k$ becomes prohibitively expensive. Furthermore, as the number of parameters $\mLam$ included in the waveform model increases in order to more accurately describe the physics of the system, the likelihood must be sampled extensively. This repeated evaluation has become a bottleneck in gravitational wave parameter estimation. \\

\subsection{ROQ rule for gravitational wave models}
\label{sec:roq_rule}

Here we describe succinctly the procedure of the ROQ rule building, following the conventions in Ref.~\citep{Smith2016}. A gravitational-wave signal model and its overlap with itself can be represented by empirical interpolants, which can be written as (cf. Eq.~(7) of Ref.~\citep{Smith2016})
%\newpage % hack to get footnote 2 to appear
%\begin{widetext}
\begin{subequations}\label{eq:EIM}
\begin{gather}
\tilde{h}_{\text{A}}(f_i;\mlam) \approx  \sum_{j=1}^{N_{\tt L}} B_j (f_i) \tilde{h}_{\text{A}}(F_j;\mlam)\,, \text{with}\, \text{A} \in \{ +,\times \}  \label{eq:EIM_lin}\,,\\
 \Re  \left[ \tilde{h}_{\text{A}}(f_i;\mlam)\tilde{h}_{\text{B}}^*(f_i;\mlam) \right]
\approx \sum_{k=1}^{N_{\tt Q}} C_k (f_i) 
     \Re \left[ \tilde{h}_{\text{A}}(\mathcal{F}_k;\mlam)\tilde{h}_{\text{B}}^*(\mathcal{F}_k;\mlam) \right]\nonumber\,,\\
\quad\quad\quad\quad\quad\quad\quad\quad\quad\quad\quad\quad\quad\text{with}\, \text{A},\text{B} \in \{+,\times\}\label{eq:EIM_quad} \,,
\end{gather}
\end{subequations}
%\end{widetext}
that accurately approximate both the polarization states and their products that are required to compute the log-likelihood in Eq.~(\ref{eq:likelihood}).
$\mlam$ is a subset of $\mLam$. It consists of the parameters that have non-trivial effects on the waveform’s amplitude and phase, such as its component masses and spin vectors \citep{Smith2016}. $\{B_j\}_{j=1}^{N_{\tt L}}$ is the reduced basis (RB) of size $N_L$ for the two polarization states and $\{C_k\}_{k=1}^{N_{\tt Q}}$ is the reduced basis of size $N_Q$ for the inner product of the waveform with itself. $\{F_j\}_{j=1}^{N_{\tt L}}$ are the \textit{empirical interpolation nodes}, which are uniquely selected to produce accurate waveform interpolation with the basis $\{B_j\}_{j=1}^{N}$.
The $\tilde{h}_{\text{A}}(\mlam; F_j)$ is an \textit{empirical
interpolant} of the $\text{A}$-polarization state at those empirical nodes.  
Similarly for the products of polarization states. 
Substituting the approximation in Eq.~(\ref{eq:EIM}) into
Eq.~(\ref{eq:inner}) generates a reduced order quadrature rule. Section~\ref{sec:algorithms} describes the algorithm we use to build the bases in Eq.~\eqref{eq:EIM}. 

%We break the likelihood into those pieces which we can approximate using~\eqref{eq:EIM}

All extrinsic parameters (defined here as the sky-position RA and DEC, the polarization angle $\psi$, the distance $D$ and the coalescence
time $t_c$), do not affect the frequency evolution of the binary and simply scale the inner product, thereby sharing the same ROQs, except for the coalescence time $t_c$.  The coalescence time does require special treatment: following previous work (see \cite{Smith2016,Canizares2014}, we build a unique set of ROQ weights for equally spaced values of $t_c$ (see below).

The full likelihood can be approximated by the ROQ likelihood, which can be decomposed into parts \citep{Smith2016}

\begin{equation}
\label{eq:ROQ_likelihood1}
\begin{aligned}
\log\mathcal{L}\approx &\ F_+ (d,h_+)_{\text{ROQ}} + F_\times (d,h_\times)_{\text{ROQ}} \\ &- F_+ F_\times (h_+,h_\times)_{\text{ROQ}} - \frac{1}{2}\left[\big|F_+\big|^2 (h_+,h_+)_{\text{ROQ}}\right.\\
&\left.+\big|F_\times\big|^2(h_\times,h_\times)_{\text{ROQ}}+(d,d)\right] \, ,
\end{aligned}
\end{equation}
where the linear part and its corresponding ROQ weights are
\begin{subequations}\label{eq:ROQ_part1}
\begin{eqnarray}
&&(d,h_{\text{A}}(\mlam) )_{\text{ROQ}}
\approx\sum_{j=1}^{N_{\tt L}}\omega_j(t_c) \tilde{h}_{\text{A}}(F_j;\mlam) \,,\\
&&\omega_j(t_{c}) = 4\Re\ \Delta f\sum_{i=1}^L \frac{ \tilde{d}^*(f_i) B_j (f_i)}{ S_n(f_i)} \mathrm{e}^{-2 \pi \mathrm{i} t_{c} f_i } \,,
\end{eqnarray}
\end{subequations}
and the quadratic part and its weights are
\begin{subequations}\label{eq:ROQ_part2}
\begin{eqnarray}
&&(h_{\text{A}}(\mlam),h_{\text{B}}(\mlam) )_{\text{ROQ}}
\approx \sum_{k=1}^{N_{\tt Q}}\psi_k \tilde{h}_{\text{A}}(\mathcal{F}_k; \mlam)\tilde{h}^{*}_{\text{B}}(\mathcal{F}_k;\mlam) \,,\\
&&\psi_k = 4\Re\ \Delta f \sum_{i=1}^L \frac{C_k (f_i)}{ S_n(f_i)}\,.
\end{eqnarray}
\end{subequations}

Once the weights are computed, evaluating the ROQ likelihood only requires $N_{\tt L}+N_{\tt Q}$ terms, hence reducing the cost in Eq.~(\ref{eq:inner}) by a factor of $L/(N_{\tt L}+N_{\tt Q})$. The ROQ rule is similar to the standard evaluation pattern, thereby allowing existing codes to easily implement these tools. 

%%%%%%%%%%%%%%%%%%%%%%%
\subsection{Requirements on reduced bases}
\label{sec:requirementsonreducedbases}

The key ingredients for applying the ROQ technique to gravitational wave inference are the reduced bases $B_j(f_i)$ and $C_k(f_i)$ in Eqs.~\ref{eq:ROQ_part1} and \ref{eq:ROQ_part2}. The search for a reduced basis aims at a set of basis elements that are most different from each other and can span the waveform space accurately. Such bases are not unique. This can be seen by comparing the waveform space with a 3-D Euclidean vector space, where there are infinite choices for its basis.
Also, note that reordering the basis elements in a basis do not change the accuracy they represent a test waveform. These make the basis search process more flexible. The task then simplifies to searching for a basis out of many possible bases that are roughly the same size and can span the waveform space accurately. 

Three aspects should be checked carefully for any code developed to construct reduced bases and reduced order quadrature data: 

$(a).$ The accuracy of the interpolated waveforms by a reduced basis. The quality of a basis is measured by the \textit{empirical interpolation error} denoted by $\sigma_{\texttt{EI}}$ in Eq.~(21) of \citep{Field2013}. The empirical interpolation error is the self complex scalar product of the difference between the interpolation of a test waveform with the basis and the original test waveform, i.e., $\sigma_{\text{EI}}= \| h( \mlam ) - \mathcal{I}_{N_L}[h(\mlam)] \|^2$ where $\mathcal{I}_{N_L}[h(\mlam)]$ denotes the empirical interpolant on the right side of Eq.~(\ref{eq:EIM_lin}) with a basis of size ${N_L}$. To ensure the waveforms represented by a basis do not introduce systematic errors due to the interpolation in a parameter estimation process \citep{canizares2013, Canizares2014}, we set the waveform interpolation errors to be one order of magnitude smaller than the accuracy of waveform model. While it is possible to build ROQs of better accuracy, it is not necessary and will lead to larger basis sizes. %On the contrary, more accurate bases will cause less speedup in parameter estimation because their sizes are larger. 

$(b).$ The sizes of the reduced linear and quadratic bases. Smaller basis sizes enable greater speedup in gravitational wave inference. The number of basis elements needed to precisely represent the waveforms in the waveform space depends on two factors, the complexity of the waveform model%, i.e., how different are the wiggles in each waveform, 
and the durations of the waveforms. When a basis size is too large, e.g., over a few thousands, the parameter ranges should be split into smaller chunks to reduce the basis size. Waveform space is usually split based on chirp mass range, which corresponds to signal duration ($T$) range. It is unnecessary to have a one-fits-all basis that can represent all the waveforms in the whole parameter space of a gravitational wave model, because the basis would be too large to save time for parameter estimation. Also, gravitational wave search pipelines such as \gstlal and \pycbc \citep{gstlal, pycbc} can estimate chirp mass at $10\%$ precision for BBH mergers and $1\%$ or even $0.1\%$ precision for BNS mergers \citep{gwtc-1,gwtc-2}, so it is sufficient to use a basis built from a small chunk of waveform space. In addition, one can run parameter estimation analyses in parallel with multiple bases and combine the results in post-processing, using the evidence from Eq. (\ref{eq:posterior}).

$(c).$ The time it takes to build the basis. In practice for LVK analyses we set a maximum of two weeks for any basis construction. Smaller chunks of parameter ranges and more computing resources are used to adjust the building time.

%%%%%%%%%%%%%%%%%%%%%%%
\subsection{Numerical algorithm of \pyroq}
\label{sec:algorithms}

In this section we describe the algorithm that \pyroq uses to search for reduced bases and build ROQ data. 

For a given waveform model, we first need to know its accuracy, its parameters, and the parameter ranges. 
The waveform accuracy determines how accurate the reduced bases need to interpolate the waveforms. The criterion we set is that the maximum empirical interpolation error, denoted by $\sigma_{\text{EI,max}}$, in the training waveform set is one order of magnitude smaller than the accuracy of the waveform model. With this criterion, most of the waveforms in the training set are represented several orders of magnitude more accurately than the waveform model's uncertainty.  We use letter $\epsilon$ to denote the threshold. 

The number of waveform parameters and the parameter ranges determine how many chunks the waveform space should be sliced into and how. The slicing of the chunks for the {\tt IMRPhenomPv2} waveform model in Table I of \citep{Smith2016} can be used as a reference. When a waveform model has more parameters and physics, its parameter ranges may have to be sliced into smaller chunks than {\tt IMRPhenomPv2}, so that the basis sizes remain practical. Another downside when a chunk is too big is that the training set of the waveforms has to be large too in order to sample the possible features of the waveform model. Larger training sets increase the computational costs of reduced basis building. Note that overlaps are made between adjacent chunks to cover the situation where a signal is near the dividing boundary.

After the chunks are sliced, they are handled separately. For any chunk of waveform space, the aim is a linear basis and a quadratic basis that can span that waveform space and the space of the waveform self-overlaps accurately. The search algorithm for a linear basis is shown in Algorithm \ref{alg:searching}. The strategy is the same for a quadratic basis, only replacing the waveforms by the overlaps of waveforms with themselves. 
The outline of the search process is as follows. The reduced order quadrature rule is trained on a large training set ($\mathcal{T}$) of waveforms of parameters $\mlam$'s, with the training starting from a small subset of the training waveforms.  After the small subset can be interpolated accurately, the training proceeds to a larger subset that includes the previous smaller subset. The training progresses several rounds, each round with a larger subset of the total training set. The training is completed after all the training waveforms can be accurately interpolated, i.e., the maximum interpolation error of the training set is smaller than the threshold.  

To start the training, initial basis elements are required. Theoretically, any waveforms in the chunk of waveform space can serve the purpose. However, the interpolation is most efficient by using the corners of the parameter space as the initial basis elements. From them the first few basis elements are extracted out correspondingly using the Gram-Schmidt process. This choice of the initial basis elements has an outsized impact on the eventual basis sizes. In practice, using the corner waveforms rather than random waveforms as initial basis elements results in an eventual basis size being reduced by up to $20\%$. The number of corner initial waveforms is denoted by $N_c$ in Algorithm~\ref{alg:searching}.  

With the few initial basis elements, we first search for new basis elements in a subset of the total training waveforms. The subset size is denoted by $N_{\text{sub}}$.
Typically a training set has 10 million waveforms. A subset is the first $N_{\text{sub}}$ training waveforms, and is denoted by $\mathcal{T}_{N_{\text{sub}}}$ in Algorithm~\ref{alg:searching}. For example, $\mathcal{T}_{1000}$ is the subset that has the first 1000 training waveforms. We have an optional pre-selection process to search for basis elements, where Gram-Schmidt projection is applied to select the most different basis elements from the $\mathcal{T}_{100000}$ subset to interpolate $\mathcal{T}_{1000}$ accurately. In Algorithm~\ref{alg:searching} this projection is denoted by $\mathcal{P}_j[h(\mlam)]$ for the projection of $h(\mlam)$ with a basis of size j. The pre-selection also enables quick learning about the complexity of the waveform model and the basis size of the chunk of waveform space.

\begin{algorithm}[H]
\caption{Search for a linear basis}
\label{alg:searching}
\begin{algorithmic}%%%%%%%%%%%%%%%%%%%
\State {\bf Input:} $\{{\bf e}_{\text{corner},j}\}^{N_c}_{j=1}$,  $\epsilon$, $\mathcal{T}$ 
\vskip 10pt
\State Set $i=0$  %and define $\sigma_{EI,0} = 1$
\State RB = $\{{\bf e}_{\text{corner},j}\}^{N_c}_{j=1}$ 
\While{$\sigma_{\text{EI,max}} \ge \epsilon$}~{(Optional pre-selection loop)}
\State {$i=i+1$}
\State $\mlam_{\text{new}} = \text{argmax}_{ h(\mlam) \in {\mathcal{T}_{100000}} } \| h(\mlam ) - \mathcal{P}_j[h(\mlam)] \|^2 $ ~(Gram-Schmidt)
\State $e_{i+1} = h(\mlam_{\text{new}} ) - \mathcal{P}_i[h(\mlam_{\text{new}})]$
\State $e_{i+1} = e_{i+1} / \| e_{i+1} \|$ 
\State RB = RB $\cup \, e_{i+1}$
%\EndFor
\State{$\sigma_{\text{EI,max}} = \max_{ \mlam \in {\mathcal{T}_{1000}}} \|h(\mlam )-\mathcal{I}_{i+1}[h( \mlam )]\|^2 $}
\EndWhile
\For{$\mathcal{T}_k \in \{\mathcal{T}_{10000}, \mathcal{T}_{100000}, \mathcal{T}_{1000000}, \mathcal{T}\times5\ \text{times}\}$
}
\While{$\mathcal{T}_{\text{outliers}} \neq \varnothing$} 
\State{$\mathcal{T}_{\text{outliers}}$ = $h(\mlam) \in \mathcal{T}_k$ and $\|h(\mlam )-\mathcal{I}_i[h(\mlam)]\|^2 > \epsilon$ }
\While{$\sigma_{\text{EI,max}} \ge \epsilon$}
\State{i=i+1}
\State{$\sigma_{\text{EI,max}} = \max_{ h(\mlam) \in {\mathcal{T}_k}} \|h( \mlam )-\mathcal{I}_i[h( \mlam )]\|^2 $}
\State $\mlam_{\text{new}} = \text{argmax}_{ h(\mlam) \in {\mathcal{T}_k} } \| h( \mlam ) - \mathcal{I}_i[h(\mlam)] \|^2 $
\State $e_{i+1} = h( \mlam_{\text{new}}) - {\cal P}_{i} h(\mlam_{\text{new}})$ 
\State $e_{i+1} = e_{i+1} / \| e_{i+1} \|$ {\hskip0.725in} 
\State RB = RB $\cup \, e_{i+1}$
\State $\mathcal{T}_{\text{outliers}} = \mathcal{T}_{\text{outliers}} - h(\mlam_{\text{new}})$
\EndWhile
\EndWhile
\EndFor
\vskip 10pt
\State {\bf Output:} RB $\{ e_i \}_{i=1}^{N_L}$ 
\end{algorithmic}
\end{algorithm}

\begin{figure}[htp]
    \centering
    \includegraphics[height=8cm]{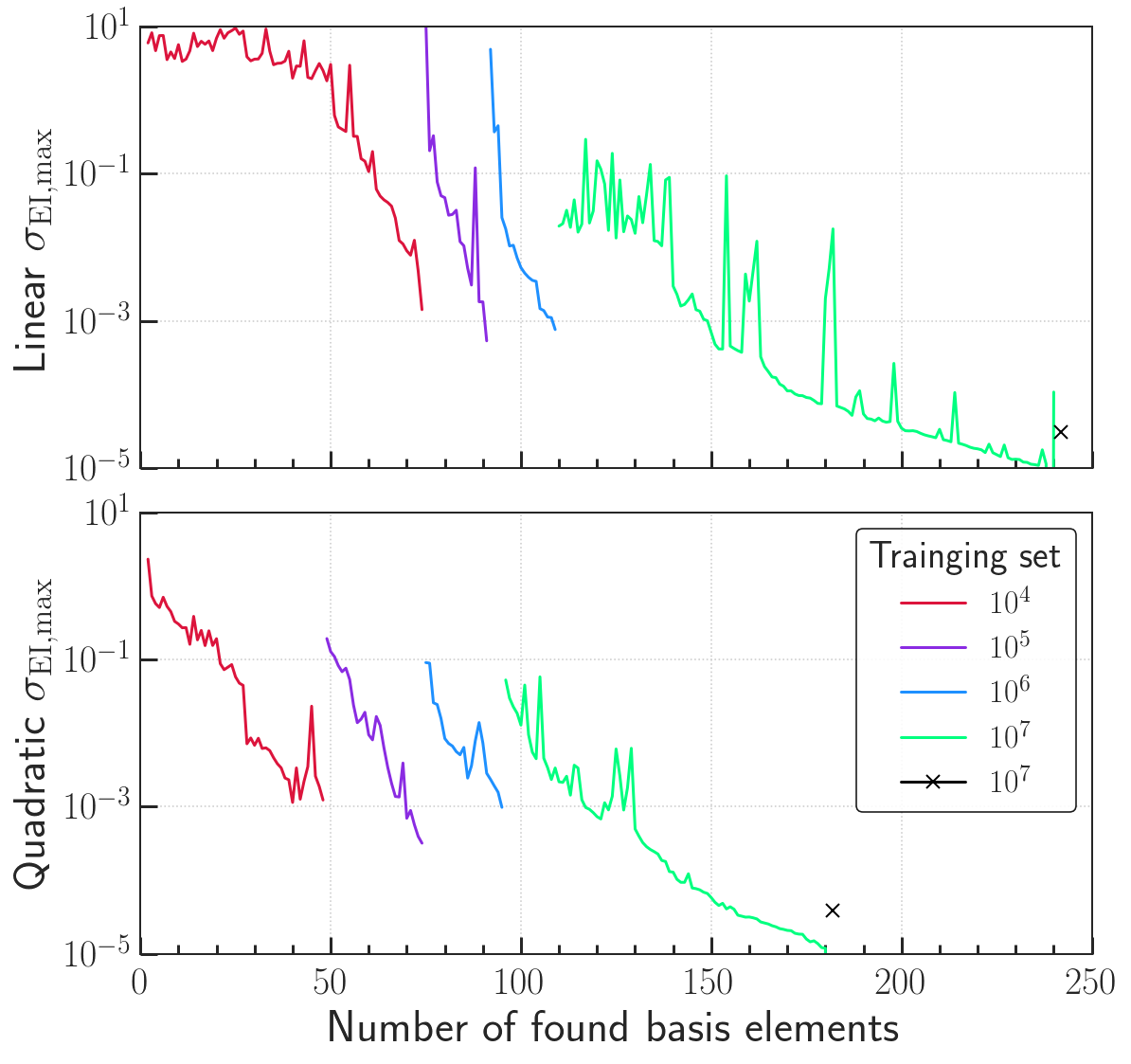}
    \caption{The maximum empirical interpolation error as a function of the number of the basis elements for the 4-second waveforms using the {\tt IMRPhenomPv2} model; also refer to Bases 4s-P in Table~\ref{tab:4secondbasescomparison}. The top panel is for the linear basis and the bottom panel is for the quadratic basis. The colors are for the different rounds of training with subsets of the total training waveforms, which are 10 millions in this case.}
    \label{fig:maxeimerrorvsbasissize}
\end{figure}

The training uses the maximum empirical interpolation error as the sole indicator for every new basis element search. This has proved to be an efficient method. The ``found basis" is applied to a small subset, $\mathcal{T}_{10000}$. In the case pre-selection is skipped, the initial corner basis is used as the found basis. Usually a fraction of the training subset cannot be represented accurately by the found basis. Those waveforms are called outlier waveforms in the training set, $\mathcal{T}_\text{outliers}$. The outlier training waveform with the largest interpolation error is added directly as a new element to the found basis. This is different from \greedycpp, where all the outlier test waveforms with empirical interpolation errors above a certain threshold are added to enrich the training waveform set. We have observed that some of the outliers with high interpolation errors above the threshold can be from the same narrow region of the waveform space, and adding one of them as a new basis element can interpolate the rest of the outliers accurately. This proves an effective and efficient way for basis building. The new found basis, which contains the previously found basis elements and the newly added one, is then applied onto the remaining outlier training waveforms. If some of them still cannot be interpolated accurately, then the one with the highest empirical interpolation error is added as another new basis element. The process is repeated until all the outlier waveforms in the subset of the training waveforms are represented accurately. Then the training moves to the next round with a larger subset of the training waveforms that includes the previous smaller training subset. The sizes of the subsets in the adjacent rounds differ by a factor of 10. In the last rounds of the training, the previously found basis is used to interpolate all the waveforms in the training set $\mathcal{T}$. If there are outlier waveforms, a new basis element is added and the process is repeated until all the waveforms can be interpolated by the found basis.

Figure~\ref{fig:maxeimerrorvsbasissize} shows the changes of the maximum empirical interpolation error as a function of the size of the found linear basis (top panel) and the found quadratic basis (bottom panel) for a 4-second waveform chunk in the training process. Refer to Bases 4s-P in Table~\ref{tab:4secondbasescomparison} for more details about the set of bases. The different colors correspond to the training subsets in the several rounds of basis search. In general, the addition of a new basis element can decrease the maximum interpolation error of the training set (or subset for the first few rounds). The found basis that can interpolate a smaller subset accurately usually cannot represent a larger subset at the same accuracy, and hence new basis elements are needed. Gravitational wave inference typically evaluates the likelihoods for 100 thousand to 1 million waveforms. The basis built from a training set of 10 million waveforms can interpolate 1 million random waveforms accurately. This will be further discussed in the next section.

Trivial parallel computation is used in \pyroq to reduce the wall time spent on waveform interpolation in the search for new basis elements. Specifically, we use the {\tt multiprocessing} package in \python. With 50 processes running at the same time, the wall time is reduced by a factor of $\sim 40$ for the {\tt IMRPhenomPv2} model. For 10 million training waveforms, we typically use 1024 parallel processes to interpolate them. As the number of the processes increases, the wall time decreases although non-linearly.

\subsection{Waveform models}
\label{sec:waveforms}
A variety of waveform models are available for compact binary coalescences (CBCs). \pyroq has included the ones that have been released in the \lalsuite software library \citep{lalsuite} and keeps up-to-date with their new released waveform approximants. However, it is easy to adapt the code to work for non-LVK waveforms, provided that the waveforms are represented in either the frequency or time domain and can be assessed at requested frequency or time nodes. For the ROQs to be used in the LVK parameter estimation pipelines, the frequency domain is in practice preferable even though there is no such requirement in ROQ data constructions.

%%%%%%%%%%%%%%%%%%%%%%%%%%%%%%%%%%%%
\section{Code Validation with {\tt IMRPhenomPv2} waveform model}
\label{sec:codevalidation}
%%%%%%%%%%%%%%%%%%%%%%%%%%%%%%%%%%%%
In this section we first  compare the performances of \pyroq with those of \greedycpp in terms of waveform interpolation accuracy and basis size. We then demonstrate that the bases generated with \pyroq can represent waveforms to the accuracy specified by a waveform model and required by parameter estimation.

The waveform model used for the code validation is the phenomenological waveform model {\tt IMRPhenomPv2}, which is implemented in the LIGO Algorithm Library (\lal) \cite{LAL}. This model describes an approximate inspiral-merger-ringdown (IMR) signal of a precessing binary black hole system by appropriately rotating the waveforms of an aligned-spin system. The waveform accuracy for this model is $1\%$ or less in the parameter ranges of interest. Theoretically, any waveform model can be used for the validation of the code, but we chose the precessing waveform model {\tt IMRPhenomPv2} for two reasons: First, it is a waveform that is frequently used in LVK data analysis. Second, the bases for this model have been built using \greedycpp, so the performances of both \greedycpp and \pyroq can be more easily and directly compared. 

\subsection{Performance compared with \greedycpp}

\pyroq was used to build linear and quadratic bases for the {\tt IMRPhenomPv2} waveform model. We compare those bases with \citep{ROQ-IMRPhenomPv2} (also refer to  \citep{Smith2016}), taken to be representative of \greedycpp as the publicly available bases referenced in the literature. The 4-second and the 8-second cases were chosen, and the parameter ranges were the same as those built by \greedycpp in \citep{ROQ-IMRPhenomPv2}. Specifically, the chirp mass $M_c$ ranges from $12.3 M_\odot$ to 45 $M_\odot$ for the 4s bases and $7.93 M_\odot$ to $14.76 M_\odot$ for the 8s bases. The magnitudes of the spin-related parameters $(\chi_1, \chi_2, \chi_p)$ lie within the range $(-0.88, -0.88, 0) \leq (\chi_1, \chi_2, \chi_p) \leq(0.88, 0.88, 0.88)$ for the 4s bases and $(-0.8, -0.8, 0) \leq (\chi_1, \chi_2, \chi_p) \leq(0.8, 0.8, 0.8)$ for the 8s bases. The spin angles are $(0,0) \leq (\theta_J , \alpha_0) \leq (\pi, 2\pi)$, and the frequency range is 20 to 1024 Hz for both the 4s and the 8s bases. The step size in frequency is $\Delta f = 1/4$ Hz for the 4s bases and $\Delta f = 1/8$ Hz for the 8s bases. 

The comparisons of the ROQ bases built by \greedycpp in \citep{ROQ-IMRPhenomPv2} and those built by \pyroq are listed in Table~\ref{tab:4secondbasescomparison}. The set of 4-second linear and quadratic bases built with \greedycpp is labeled Bases 4s-G, and the set with \pyroq is labeled Bases 4s-P. The same notation applies to the 8-second bases. Bases 4s-P and Bases 8s-P both can represent a random set of 10 million training waveforms at $10^{-4}$ or higher accuracy. The wall time of construction for Bases 4s-P by \pyroq was 70 hours for the linear basis and 57 hours for the quadratic basis, with 40 CPUs and 102 GB RAM on a LIGO Data Grid cluster. For Bases 8s-P, these numbers are 37 hours and 45 hours, with 40 CPUs and 110 GB RAM. For each basis construction, over $90\%$ of the time was spent on the calculations of interpolation errors of the training set of 10 million waveforms. 

We first compare Bases 4s-G with Bases 4s-P. With nearly half the number  ($240\ \text{v.s.}\ 464$) of linear basis elements and considerably fewer ($181\ \text{v.s.}\ 280$) quadratic basis elements, Bases 4s-P can represent a random set of one million test waveforms at better accuracy than Bases 4s-G. The linear basis in Bases 4s-P produces smaller maximum linear empirical interpolation error for the one million test waveforms, i.e., $6.4\times10^{-3}$ with Bases 4s-G and $4.1\times 10^{-3}$ with Bases 4s-P. The quadratic basis in Bases 4s-P produces smaller maximum quadratic empirical interpolation error, i.e., $1.3\times10^{-2}$ with Bases 4s-G and $4.3\times10^{-5}$ with Bases 4s-P. Bases 4s-P also has fewer waveforms of decreased interpolation accuracy for the one million test waveforms; see Table~\ref{tab:4secondbasescomparison}. The number of waveforms with linear empirical interpolation errors larger than $10^{-5}$ is 187 with Bases 4s-G and 35 with Bases 4s-P. For the quadratic empirical interpolation errors, these numbers are 732 with Bases 4s-G and 4 with Bases 4s-P. Compared to \greedycpp, \pyroq can further speed up the calculation of the likelihood function by $(464+280)/(240+181)=1.77$ times when analyzing a 4-second signal using ROQ method with Bases 4s-P, even at slightly better waveform interpolation accuracy. Similar conclusions can be drawn from the comparison for the 8s case, where the further speedup factor is 1.60 by \pyroq compared to \greedycpp while the same waveform interpolation accuracy is maintained. The conclusion is valid for those two randomly selected cases. It is important to note that we used \citep{ROQ-IMRPhenomPv2} as representative basis sets created by \greedycpp. While useful for this comparison, it is possible that further improvement on these bases could be done with \greedycpp. Whether \pyroq can build smaller-sized bases than \greedycpp for any waveform space is to be further tested. 

%\begin{widetext}
\begin{table*}[htp]
\centering
\begin{tabular}{c | c | c | c | c | c | c | c} 
\hline
\hline
\rule{0pt}{20pt}Bases & Code &  %\shortstack{$M_c (M_\odot)$ \\ Min\ \ Max} & 
\shortstack{Basis size \\ Linear\ \ Quadratic} &  \shortstack{Training $\sigma_{\text{EI,max}}$  \\ Linear\ \ Quadratic}& \shortstack{$\#$ of training\\waveforms} & \shortstack{Testing $\sigma_{\text{EI,max}}$ \\ Linear\ \ Quadratic} & \shortstack{Testing $\sigma_{\text{EI}}>10^{-5}$ \\Linear\ \ Quadratic} & \shortstack{$\#$ of test\\waveforms} \\
\hline
\rule{0pt}{13pt}4s-G & \greedycpp  & 464 \ \ \ \ 280 &---\ \ \ \ \ \ \ \ \ \ ---&\sqz$1.5\times 10^7$ &\sqz $6.4\times10^{-3}$ \ \ $1.3\times10^{-2}$ &187\ \ \ \ \ \ 732 & $10^6$\\ 
\hline
\rule{0pt}{13pt}4s-P & \pyroq & 240 \ \ \ \ 181 &\sqz $3.1\times10^{-5}$\ \ $3.9\times10^{-5}$&\sqz$1.0\times10^7$ &\sqz $4.1\times10^{-3}$ \ \ $4.3\times 10^{-5}$ & 35\ \ \ \ \ \ \ \ \ 4 & $10^6$\\
%\hline
%\rule{0pt}{13pt}4sP2 & PyROQ & 450 \ \ \ \ 308 &\sqz $5.8\times10^{-7}$\ \  $8.7\times10^{-7}$ & $ 10^7$ &\sqz $5.5\times 10^{-5}$ \ \ $6.3\times10^{-7}$  & 9\ \ \ \ \ \ \ \ \ \ 0 & $10^6$\\
\hline
\rule{0pt}{13pt}8s-G & \greedycpp  & 386 \ \ \ \ 270 &---\ \ \ \ \ \ \ \ \ \ ---&\sqz$1.5\times 10^7$ &\sqz $6.2\times10^{-2}$ \ \ $9.7\times10^{-2}$ & 479\ \ \ \ \ \ 823 & $10^6$\\ 
\hline
\rule{0pt}{13pt}8s-P & \pyroq & 238 \ \ \ \ 171 &\sqz $9.6\times10^{-4}$\ \  $8.6\times10^{-4}$ &\sqz$1.0\times10^7$ &\sqz $2.0\times 10^{-3}$ \ \ $5.3\times10^{-4}$  & 105\ \ \ \ \ \ \ 211 & $10^6$\\
%\hline
%\rule{0pt}{13pt}8sP2 & PyROQ & 375 \ \ \ \ 276 &\sqz $1.1\times10^{-5}$\ \  $9.6\times10^{-5}$ & $ 10^7$ &\sqz $9.9\times 10^{-3}$ \ \ $tbd\times10^{-3}$  & 917\ \ \ \ \ \ \ \ tbd & $10^6$\\
\hline
\hline
\end{tabular}
\caption{Comparisons of \greedycpp and \pyroq in building ROQ data with the {\tt IMRPhenomPv2} waveform model. Two sets (4s-P and 8s-P) of linear and quadratic bases were built with \pyroq in the same parameter ranges that the two sets (4s-G and 8s-P) of \greedycpp bases were constructed. Here 4s-G stands for the bases built with \greedycpp for the waveforms of around 4-second durations or the chirp mass $M_c$ ranging from $12.3 M_\odot$ to 45 $M_\odot$; the other basis sets in the table use the same naming notation. The parameter ranges can be found in \citep{ROQ-IMRPhenomPv2} (also see \citep{Smith2016}). For the two example chunks, \pyroq can interpolate test waveforms more accurately with less basis elements than \greedycpp.}
\label{tab:4secondbasescomparison}
\end{table*}
%\end{widetext}

\subsection{Accuracy of interpolated waveforms}

Parameter estimation places requirements on the accuracy of the interpolated waveforms for the ROQ method to get the same inference results as the standard method without introducing any systematic errors. The empirical interpolation errors of the waveforms should be preferably one order of magnitude smaller than the accuracy of the waveform model, such that the fractional error in logarithmic likelihood is below the errors due to waveform accuracy in the posteriors. 
\begin{figure}[htp]
    \centering
    \includegraphics[trim=0.25cm 0.25cm 0.2cm 0.2cm, clip=true,height=10.5cm]{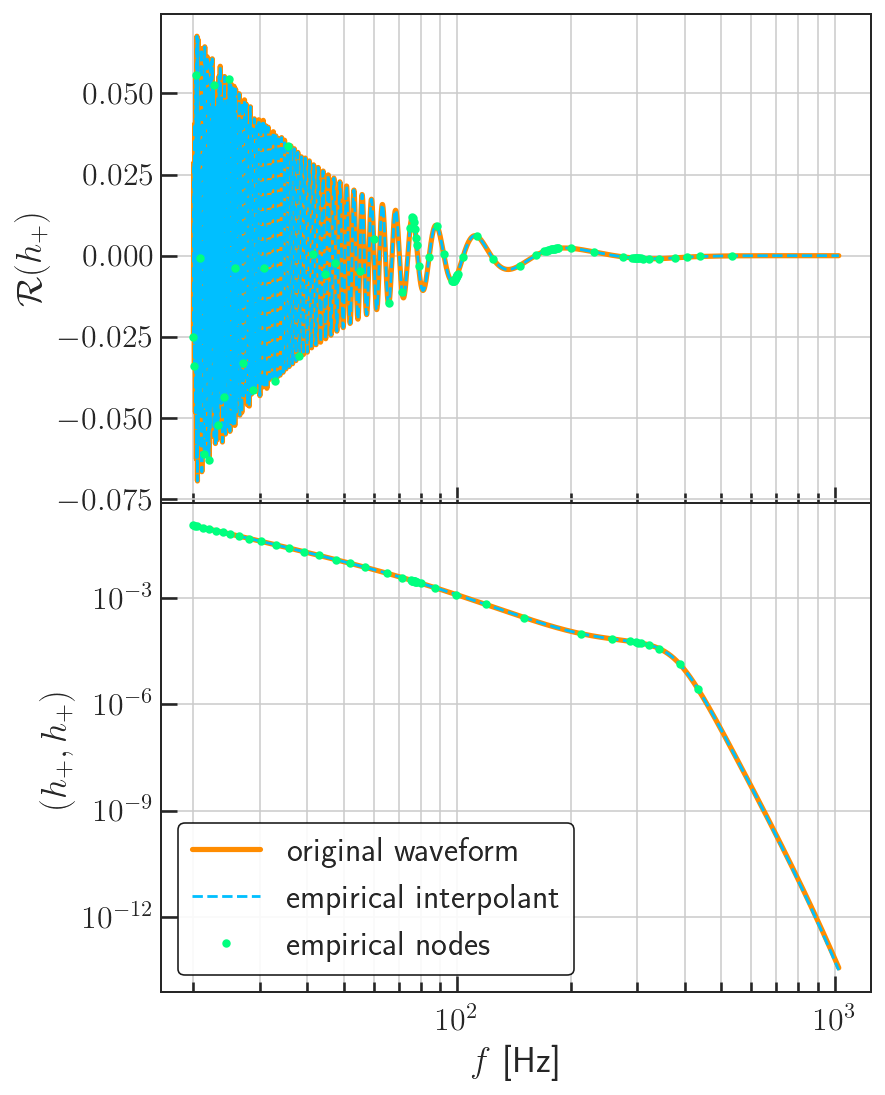}
    \caption{Empirical interpolants of a normalized random 16-second test waveform (top panel) and its overlap with itself (bottom panel).  The orange thick lines are the original waveform (top) and the original overlap (bottom). Their corresponding interpolants are shown in blue dashed lines.  The green dots are the empirical frequency nodes, the number of which is equal to that of the reduced basis elements that are used to represent the waveform or the overlap. The empirical interpolation errors are $2.1\times 10^{-13}$ and $1.7\times 10^{-14}$, respectively. The test waveform is from an NSBH merger with $\mathcal{M}_c=7.0\ M_{\odot}$ and $q=14.0$, using the model {\tt IMRPhenomPv2}. 
    }
    \label{fig:empiricalinterpolant}
\end{figure}

\begin{figure}[htp]
    \centering
    \includegraphics[trim=0.25cm 0.25cm 0.2cm 0.2cm, clip=true,height=5.5cm]{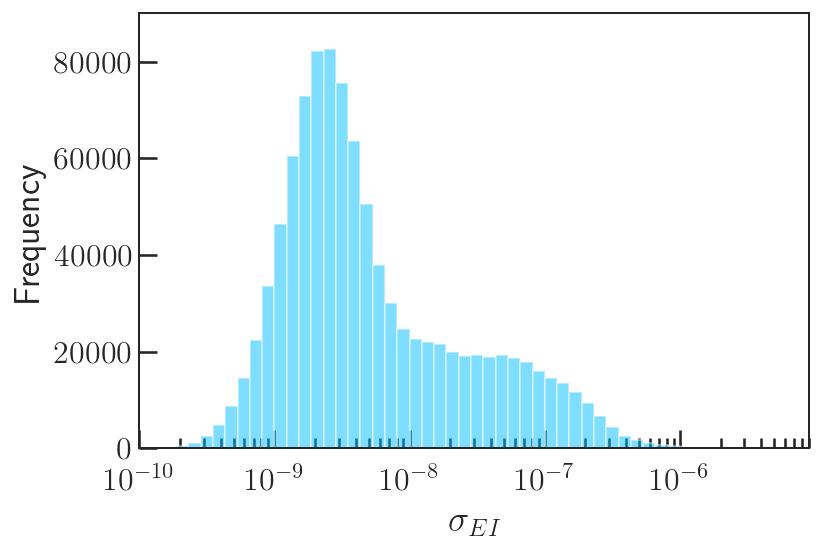}
    \caption{The distribution of the empirical interpolation errors of one million randomly drawn samples from the parameter ranges that are used in the construction of the 4-second linear basis in Bases 4s-P in Table~\ref{tab:4secondbasescomparison}. All test samples have interpolation errors less than $4.1\times 10^{-3}$ and greater than $10^{-12}$. The empirical interpolation errors larger than $10^{-5}$ or smaller than $10^{-10}$ are not plotted in the histogram because they are too few (35 and 73, respectively) to show in the figure.}
    \label{fig:waveformerrors}
\end{figure}

Figure~\ref{fig:empiricalinterpolant} demonstrates an instance of the linear and the quadratic empirical interpolants of a random test waveform and its overlap with itself. The original waveform is generated by \lalsimulation, a package of \lalsuite. %On the top panel of Fig.~\ref{fig:empiricalinterpolant} the original waveform is shown in orange, the interpolated waveform is shown in blue, and the empirical frequency nodes are shown in green. In the bottom panel the orange, blue, and green color are the overlap of the waveform with itself, its empirical interpolation, and the corresponding empirical nodes. 
%A 4-second waveform with Bases 4s-P could have been used in the instance, but here we choose a longer, 16-second waveform from an NSBH merger because it has more features and shows more clearly the capability of the ROQ technique. We also use this waveform later in the paper to demonstrate the parameter estimation of an injection with it and the significant speedup. 
The waveform has chirp mass $\mathcal{M}_c=7 M_{\odot}$ and mass ratio $q=14$. The waveform and its overlap with itself are interpolated with a set of bases built from a chunk of 16s waveforms. The sizes are 85 and 45 for the linear basis and the quadratic basis.

We also randomly drew one million waveform samples and calculated their empirical interpolation errors using the linear basis of Bases 4s-P in Table~\ref{tab:4secondbasescomparison} to check if the accuracy requirement can be satisfied. Note that the waveform accuracy of the model {\tt IMRPhenomPv2} is 0.01. Therefore the accuracy of the test waveforms interpolated by the basis is required to be at better than 0.01 level and preferably 0.001 to not introduce systematic errors in parameter estimation. The empirical interpolation errors of those one million test waveforms are all smaller than $10^{-3}$, with only one exception $4.1\times 10^{-3}$, still better than the accuracy of the waveform model. Their distribution is shown in Fig.~\ref{fig:waveformerrors}. The majority of the waveform interpolation errors are between $10^{-10}$ to $10^{-6}$, which are far more accurate than the waveform model itself.% and guarantee no systematic errors from the waveform interpolation.  

%%%%%%%%%%%%%%%%%%%%%%%%%%%%%%%%%%%%
\section{Applications on Parameter Estimation}
\label{sec:applications}
%%%%%%%%%%%%%%%%%%%%%%%%%%%%%%%%%%%%
With the waveform bases generated by \pyroq, we study if they can be reliably applied to the inference of gravitational waves. The detectors of interest include, but are not limited to, the ground-based LIGO-Virgo-KAGRA detectors and the LISA space detector.  Simulated gravitational wave injections are used first to compare the parameter estimation posterior results between the standard method and the ROQ method. Observed detections contain noise and thus have more variables that can prevent from pinpointing the effects from waveform interpolation, but one example is given to showcase the comparison of the inferences with the standard and the ROQ method.  

\subsection{Simulated gravitational wave injections}
\label{subsec:injections}
Simulated gravitational wave injections are used to investigate the effects of empirical interpolation errors on parameter estimation. %For the questions under consideration, any type of gravitational waves from the compact binary coalescence category can fulfill the purposes. 
Here we illustrate the parameter estimation analysis with neutron star-black hole mergers, which are the latest type of CBCs detected by LIGO and Virgo \citep{S200105ae-gcn, S200115j-gcn, nsbh2020jan-followup}. Simulated signals modeled by the {\tt IMRPhenomPv2} waveform model and randomly drawn from the parameter space of the 16-second signals were injected coherently into the two LIGO detectors. 

Each simulated signal was injected using the ``zero noise approximation". Further studies are planned including the LIGO design noise and the expected noise levels in future observing runs such as O4 \citep{Abbott:2020qfu}. 
Injections made with zero noise remove biases from noise realisation and instead focus on potential biases from waveform interpolation. These injections eliminate the unclear influences on likelihood calculations from the noisy observed data, and thus reveal the possible differences resulted from the interpolations of the waveforms by the reduced bases. %The O4 noise and the LIGO design noise cases make it possible to see the effects that come from the noise.
Particularly, the study with one randomly chosen waveform is elaborately explained in Secs.~\ref{subsec:likelihoodcomparison} and \ref{subsec:posteriorcomparison}.

\subsection{Point-by-point likelihood comparisons}
\label{subsec:likelihoodcomparison}
As shown in Sec.~\ref{sec:codevalidation}, with the bases constructed by \pyroq, the errors of the empirically interpolated {\tt IMRPhenomPv2} waveforms are very small. Now we determine how those interpolation errors affect parameter estimation using the NSBH injections that are described in Sec.~\ref{subsec:injections}. Before performing parameter estimation and checking the posteriors, we first examine the likelihoods. We perform point-by-point comparisons between the likelihoods calculated with both the standard full likelihood function and the ROQ likelihood function. The likelihoods are evaluated with \bilby, a gravitational wave inference tool used in the LIGO-Virgo-KAGRA Collaboration. 

\begin{figure}[htp]
\centering
\includegraphics[trim=0.25cm 0.2cm 0.2cm 0.2cm, clip=true, height=10.5cm]{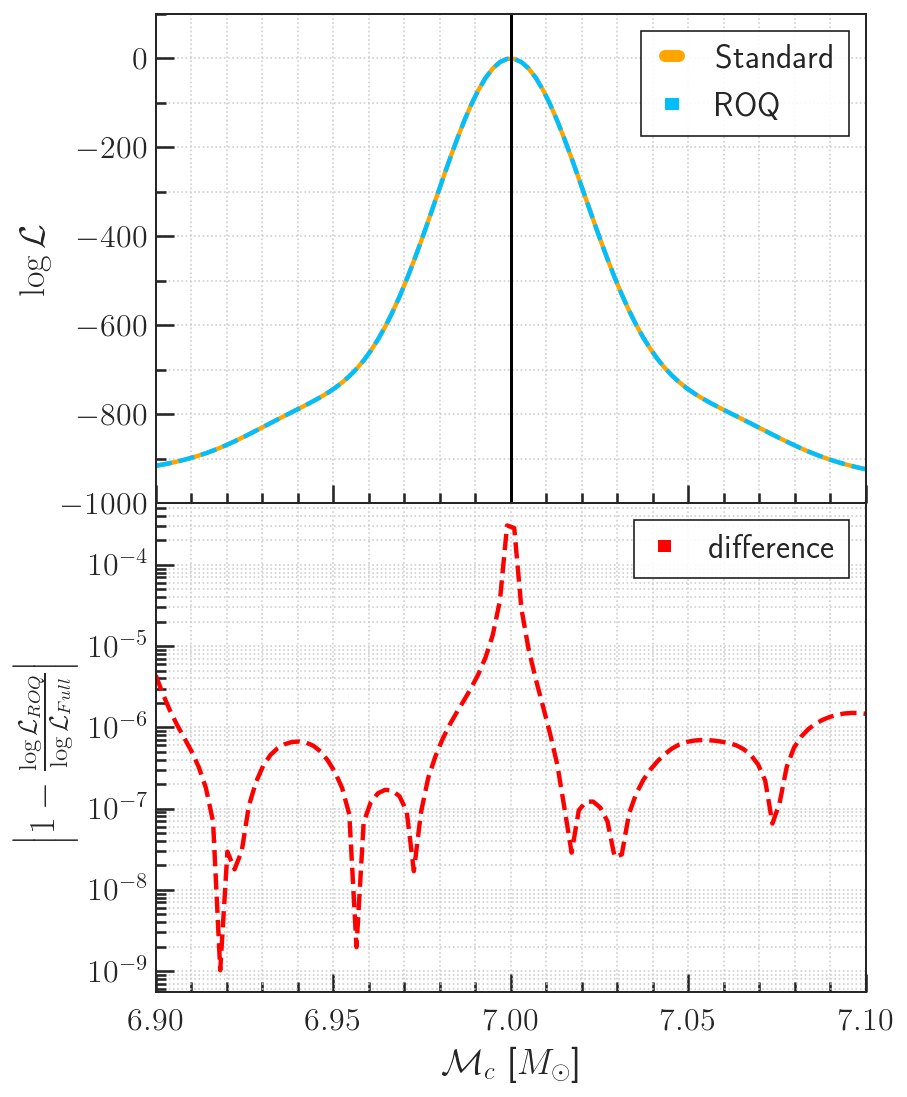}
\caption{Point-by-point likelihood comparisons between the standard full likelihood and the ROQ likelihood for a synthetic NSBH signal in zero noise using \bilby. 
The vertical black line shows the injected chirp mass value. }
\label{fig:likelihoodcomparison}
\end{figure}

Under the assumption that the ROQ is an approximation of the full likelihood, the two methods should yield the same likelihood for the same set of waveform parameter values for the ROQ to qualify a valid substitute to the full likelihood function for parameter estimation.  Figure~\ref{fig:likelihoodcomparison} shows the full and the ROQ likelihoods calculated for a randomly chosen NSBH waveform that was injected into zero noise. The injected signal has chirp mass $\mathcal{M}_c=7.0 M_{\odot}$, mass ratio $q=14.0$, luminosity distance $d_L=100$ Mpc, and the spins of both the black hole and the neutron star are 0.1. The shape of the waveform is shown in Fig.~\ref{fig:empiricalinterpolant}, where the waveform is normalized. In this likelihood comparison example the chirp mass parameter is varied and the other parameters are fixed for the likelihood calculations and comparisons, but it can be easily generalized to cases where all the parameters vary simultaneously. 
It can be seen that the likelihoods from the two methods are indistinguishable, up to a very small fractional difference
\begin{equation}
\left| 1-\frac{\log \mathcal{L}_{\texttt{ROQ}}}{\log \mathcal{L}_{\texttt{Full}}} \right| \leq 3\times10^{-4}.  
\end{equation}
%This shows that both the waveforms and the overlaps of the waveforms with themselves interpolated by the bases generated by \pyroq can represent the original ones accurately enough such that the accuracy of the likelihoods required by the parameter estimation is satisfied. 
Aside from the example case shown here, we have observed this small fractional difference to be true in the likelihood comparison of all other injection cases. %This demonstrates that \pyroq can build qualified ROQ data for the parameter estimation of gravitational wave detections where the detector noise is zero. 
Furthermore, because the ROQs were built without any noise, they can also be used in the data analysis of detections with any detector sensitivity. 
For example, a comprehensive study of the effects of the detector design noise on gravitational wave parameter estimation using the \pyroq built ROQs for the {\tt IMRPhenomNSBH} waveform model will be presented in \citep{nsbhscenario2021}.

\subsection{Standard and ROQ posterior comparisons for synthetic gravitational waves}
\label{subsec:posteriorcomparison}
%In the previous section, the ROQ likelihood is demonstrated a quality substitute for the full likelihood through point-by-point likelihood comparisons. 
In this section we compare the posterior probability distributions for the zero-noise NSBH injections discussed in Secs. \ref{subsec:injections} and \ref{subsec:likelihoodcomparison}. The parameter estimations of these synthetic gravitational waves were performed using both the standard full likelihood method and the ROQ method with the \bilby inference tool.

\begin{table}[htp]
\centering
\begin{tabular}{ c | c | c | c } 
\hline
\hline
\rule{0pt}{12pt}Parameter & Injection & Standard & ROQ  \\
\hline
\rule{0pt}{12pt}$M_c  [M_{\odot}]$ & 7.0000 & $6.9999^{0.0016}_{0.0016}$ &$6.9999^{0.0017}_{0.0014}$\\
\hline 
\rule{0pt}{12pt}q & 14.0000 & $14.0004^{+0.0079}_{-0.0089}$ & $14.0003^{+0.0083}_{-0.0082}$ \\ 
\hline 
\rule{0pt}{12pt} SNR & $33.40$ & $33.40$ & $33.40$ \\
\hline
\hline
\end{tabular}
\caption{The injected and recovered values of chirp mass $M_c$, mass ratio $q$, and signal-to-noise ratio for the example NSBH injection, with the posterior plots shown in Fig.~\ref{fig:pecomparisonfornsbhinjection}. The modes and the $68\%$ credible intervals are read from the posterior distributions produced by the standard and the ROQ methods, respectively.
}
\label{tab:injection}
\end{table}

\begin{figure}[htp]
    \centering
    \includegraphics[trim=0.4cm 0.2cm 0.2cm 0.2cm, clip=true,height=9cm]{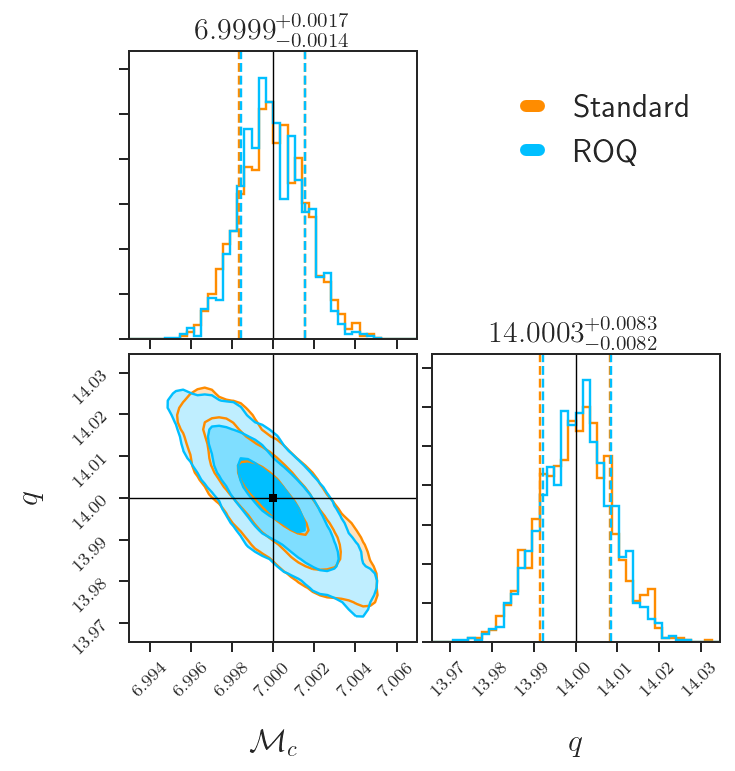}
    \caption{Corner plots of the probability density functions for the chirp mass $M_c$ and  mass ratio $q$ of a simulated zero-noise NSBH injection. 
    In orange it was obtained in about $50$ hours in terms of wall time with the standard full likelihood method, and in blue it was obtained in about 2.6 hours with the ROQ method. The injected values are shown in black vertical thin lines in each subplot, and the recovered values for the ROQ run are shown on the top of the subplots. The comparisons of the standard and the ROQ posteriors are listed in Table~\ref{tab:injection}. The inference pipeline was \bilby.}
    \label{fig:pecomparisonfornsbhinjection}
\end{figure}
The example injection discussed in detail on the likelihood comparison in Sec. \ref{subsec:likelihoodcomparison} is used to illustrate the posterior comparison in this section. 
Table~\ref{tab:injection} lists the injected and the recovered parameter values with the standard and the ROQ methods. Both methods recovered the injected gravitational wave parameters accurately and precisely. In addition, they agree with each other to the fifth significant figure on the recovered modes of the parameters, i.e., both methods have  $M_c=6.9999 M_{\odot}+\mathcal{O}(0.0001) M_{\odot}$ and $q=14.000+\mathcal{O}(0.001)$. The parameter uncertainties from the two methods differ by less than $4\%$ for all the recovered parameters in the $68\%$ credible intervals. Figure~\ref{fig:pecomparisonfornsbhinjection} shows the corner plots of the posterior distributions of the chirp mass $M_c$ and the mass ratio $q$ of the injection from the two methods. The corresponding posterior distributions are almost identical for the two methods. The slight difference is due to the sampler, i.e., we will see similar tiny posterior differences when running the parameter estimation twice with the standard method.

While the posteriors from the two methods are visually identical, the run time is different for the two methods. The ROQ method enables a 20-times speedup for the 16-second NSBH signal injection in Table~\ref{tab:injection} and Fig.~\ref{fig:pecomparisonfornsbhinjection}. Similar speedup and posterior consistency also showed in the recoveries of tens of other 16-second NSBH synthetic signals using the two methods. Note that the speedup depends on the sizes of the reduced bases compared to the frequency nodes that are determined by the signal duration, as well as the other parts of parameter estimation. The more the parameter estimation is dominated by waveform generation, the greater speedup the ROQ method can produce.

\subsection{Consistency comparison for GW150914}

%It is important to examine whether the parameter estimation results for an observed detection are consistent using the standard full likelihood method and the ROQ method with \pyroq built bases. 

\begin{figure}[ht!]
    \centering
    \includegraphics[trim=0.25cm 0.2cm 0.2cm 0.2cm, clip=true,height=6.2cm]{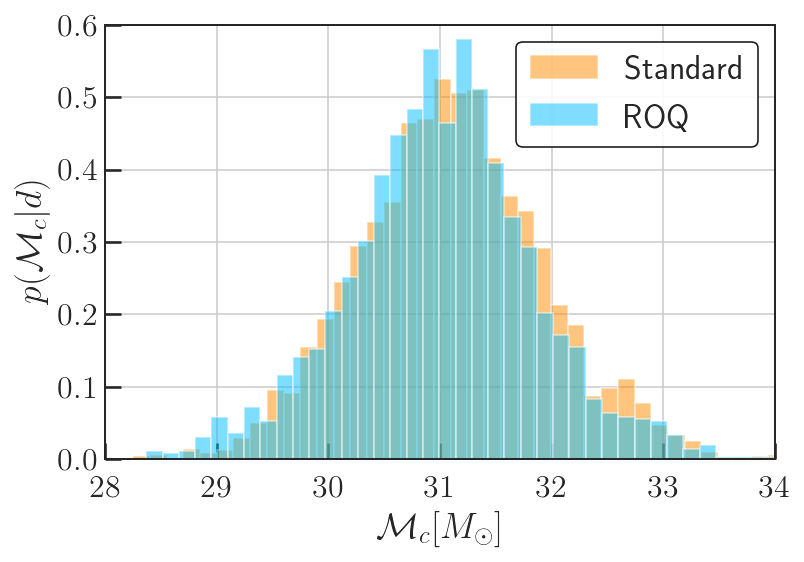}
    \caption{Comparisons of posterior probability distributions of detector frame chirp mass $\mathcal{M}_c$ from PE runs for GW150914 using the standard full likelihood method (orange) and the ROQ method (blue) with the ROQ Bases 4s-P as described in Table~\ref{tab:4secondbasescomparison}. The inference pipeline was \lalinference.
    }
    \label{fig:gw150914comparisons}
\end{figure}

As a further comparison, we analyzed the observed data that contain the gravitational wave event GW150914 in the LIGO's first observing run using \lalinference \citep{LALInference}. Two identical parameter estimation runs using the {\tt IMRPhenomPv2} waveform model were set up, except that one used the standard full likelihood method and the other used Bases 4s-P built using \pyroq in Table~\ref{tab:4secondbasescomparison} for the ROQ likelihood function evaluation. We made comparison plots between the recovered posterior distributions of the chirp mass in the detector frame in Fig.~\ref{fig:gw150914comparisons}. 
It shows that both the ROQ method with the \pyroq built reduced bases and the standard full likelihood method recovered visually identical posteriors for the detector frame chirp mass, up to a small difference that was due to the parameter estimation sampler as discussed in Sec.~\ref{subsec:posteriorcomparison}. The other parameters show similar agreement. The consistency of the inferred astrophysical parameters from both kinds of parameter estimation runs further proves \pyroq a quality ROQ building code.

%The analyses of other LVK detections using the ROQs generated by \pyroq are out of the scope of this paper. When time allows it is useful to revisit those old detections with the new tool because it can provide more bench marking that benefits inference on future LVK gravitational wave detections using \pyroq.

\subsection{Building {\tt IMRPhenomXPHM} bases for GWs of subdominant harmonic modes and precessing spins}

As a further concrete example of constructing ROQs with \pyroq, we built the bases for the waveform model {\tt IMRPhenomXPHM} \citep{Pratten:2020ceb}, which is one of the most advanced existing models for gravitational waves from binary black hole mergers. {\tt IMRPhenomXPHM} is a phenomenological frequency-domain model for the gravitational-wave signals emitted by quasi-circular precessing binary black holes, which incorporates multipoles beyond the dominant
quadrupole in the precessing frame. The accuracy of this model is not more than 0.001 for most regions of the waveform space, with a small region reaching two orders of magnitude better accuracy. Mathematical computing wise, both the {\tt IMRPhenomXPHM} and the {\tt IMRPhenomPv2} models have eight intrinsic parameters, i.e., chirp mass, mass ratio, and six spin components.  The likelihood evaluation time for a 4-second or a 8-second {\tt IMRPhenomXPHM} waveform is on average more than 5 times that for a {\tt IMRPhenomPv2} waveform. Therefore, it is important to build ROQs for this model to make it a productive tool for gravitational wave astronomy in the
era of greatly increased detections. 

Table~\ref{tab:roq-imrphenomxphm} lists the information about the bases for the {\tt IMRPhenomXPHM} model. The parameter ranges for the first three sets of ROQs are the same as those of the precessing numerical relativity (NR) waveforms that the mismatches for {\tt IMRPhenomXPHM} were computed against: the chirp mass is $26\ M\odot\leq M_c \leq 110\ M\odot$, the mass ratio is from 1 to 4, and the spins are $(-0.8, -0.8, 0) \leq (\chi_1, \chi_2, \chi_p) \leq(0.8, 0.8, 0.8)$. The step size $\Delta f$ is usually the inverse of the maximum signal duration in the chunk of waveform space the reduced bases are built for. However, $\Delta f=1/4$ Hz was used to construct the first three sets of bases although the waveform durations are much less than 4 seconds. This is because currently in the LVK inference pipelines the minimum amount of data to analyze is 4 seconds even when the signal contained in the data is far less than 4 seconds long, which is constrained by the PSD calculation where tapering is applied at the both ends of a 4-second data chunk. Outside the parameter ranges that the waveform model have been checked against the NR waveforms, it is considered extrapolation and the waveform has less accuracy.  Therefore, we lowered the interpolation accuracy requirement by one order of magnitude to build the reduced bases for the waveforms in the extrapolation ranges. In Table~\ref{tab:roq-imrphenomxphm} we list the ROQs for the two 4-second and the two 8-second chunks. The mass ratio range is $1\leq q \leq 4$ and the spins are $(-0.8, -0.8, 0) \leq (\chi_1, \chi_2, \chi_p) \leq(0.8, 0.8, 0.8)$ for all these chunks.

\begin{figure}[htp]
    \centering
    \includegraphics[trim=0.25cm 0.25cm 0.2cm 0.2cm, clip=true,height=14.5cm]{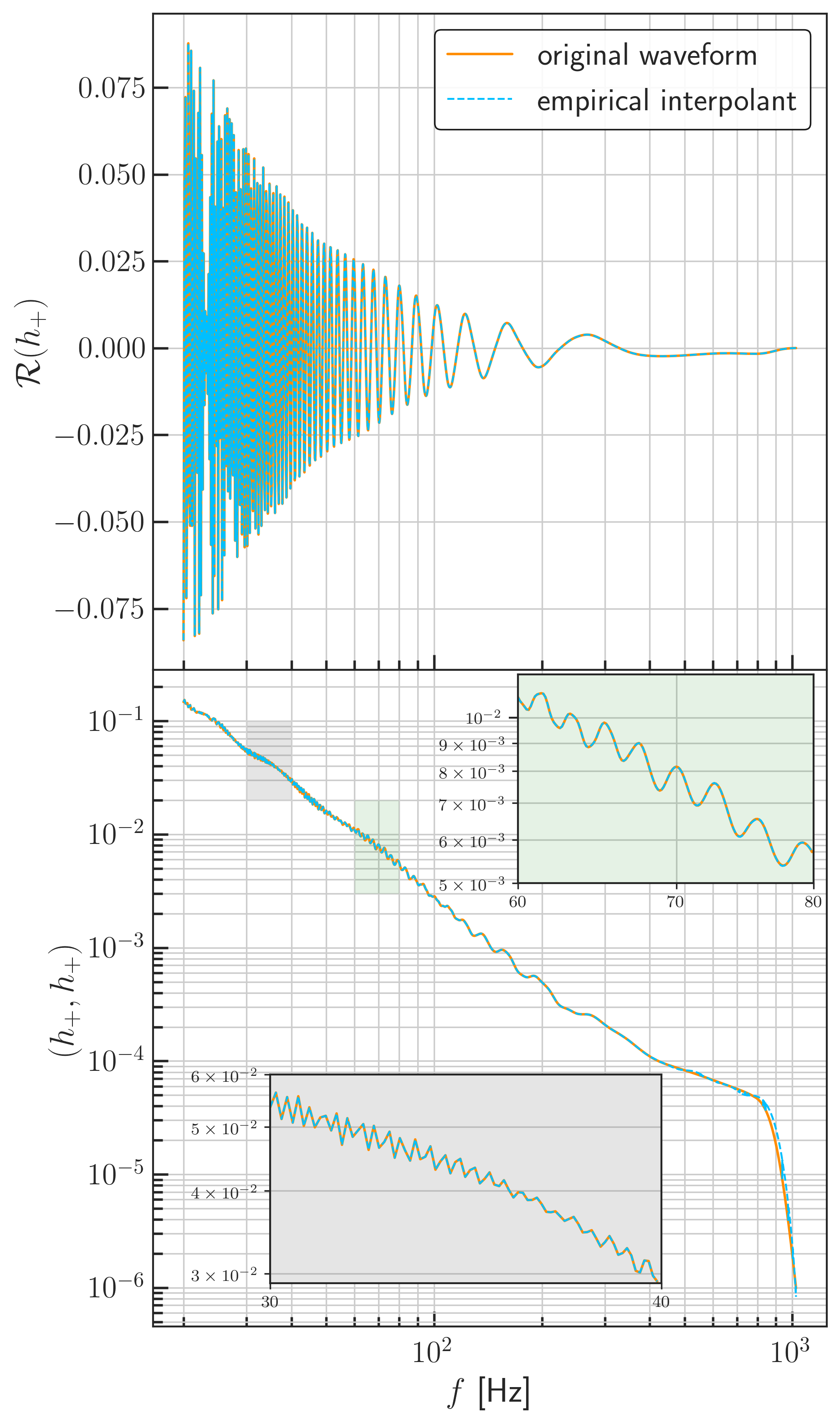}
    \caption{Empirical interpolants of a normalized 8-second test waveform (top panel) and its overlap with itself (bottom panel) with the {\tt IMRPhenomXPHM} model and the bases in Table~\ref{tab:roq-imrphenomxphm}.  The orange lines are the original waveform (top) and the original self-overlap (bottom), and the thin blue dashed lines are  the corresponding interpolants. The test waveform has $\mathcal{M}_c=8.8\ M_{\odot}$ and $q=1.8$. The empirical interpolation errors are $9.8\times 10^{-8}$ and $5.3\times 10^{-8}$, respectively. The insets in light gray and light green in the bottom panel present zoomed-in views of the features of the original and the interpolated waveform self-overlaps.  
    }
    \label{fig:xphmempiricalinterpolant}
\end{figure}

In sum, the chirp mass range for which we have built ROQs is from $8.6\ M_{\odot}$ to $110\ M_{\odot}$. The effective range for parameter estimation with the data collected by LIGO, Virgo and KAGRA detectors is about $9\ M_{\odot}$ to $100\ M_{\odot}$, which covers almost all the detected BBHs listed in GWTC-1 and GWTC-2 \citep{gwtc-1, gwtc-2}. The ROQs for the other parameter ranges will be constructed by the LVK collaboration-wide effort on ROQ building. 

\begin{table*}[ht]
\centering
\begin{tabular}{ c | c | c | c | c | c | c | c} 
\hline
\hline
\rule{0pt}{22pt} \shortstack{Freq. range (Hz) \\ Min\ \  Max}  
&  $\Delta{f} (Hz)$ &  \shortstack{$M_c (M_\odot)$ \\ Min\ \ Max}  & \shortstack{Basis size \\ Linear\ \ Quadratic} &  \shortstack{Training set $\sigma_{\text{EI,max}}$\\Linear Quadratic} & Training set &  \shortstack{Test set $\sigma_{\text{EI}}>10^{-4}$\\Linear\ \ Quadratic}& Test set \\
%\hline
%\rule{0pt}{12pt}$20\ \ \ \ \ 512$ & $1/2$ & $30\ \ 100$ & 461\ \ \ \ 272 &\sqz$5.2\times 10^{-5}$\ \ \ $5.6\times 10^{-5}$ & $1\times10^7$ &$33^*$\ \ \ \ $10^*$ & $10^6$\\
\hline
\rule{0pt}{12pt}$20\ \ \ \ 1024$ & $1/4$ & $55\ \ 110$ & 562\ \ \ \ 326 &\sqz$9.0\times 10^{-5}$\ \ \ $6.7\times 10^{-5}$ & $5\times10^6$ &$52^*$\ \ \ \ $32^*$ & $10^6$\\
\hline
\rule{0pt}{12pt}$20\ \ \ \ 1024$ & $1/4$ & $35\ \ 66$ & 731\ \ \ \ 367 &\sqz$5.5\times 10^{-5}$\ \ \ $4.7\times 10^{-5}$ & $5\times10^6$ &$152^*$\ \ \ \ $36^*$ & $10^6$\\
\hline
\rule{0pt}{12pt}$20\ \ \ \ 1024$ & $1/4$ & $26\ \ 42$ & 779\ \ \ \ 314 &\sqz$8.0\times 10^{-5}$\ \ \ $8.5\times 10^{-5}$ & $5\times10^6$ &$245^*$\ \ \ \ $57^*$ & $10^6$\\
\hline
\rule{0pt}{12pt}$20\ \ \ \ 1024$ & $1/4$ & $18\ \ 33$ & 407\ \ \ \ 214 &\sqz8.0$\times 10^{-4}$\ \ \ $8.0\times 10^{-4}$ & $5\times10^6$ &89\ \ \ \ \ \ \ 18& $10^6$\\
\hline
\rule{0pt}{12pt}$20\ \ \ \ 1024$ & $1/4$ & $12\ \ 20$ &397\ \ \ \ 266&\sqz  $6.9\times 10^{-4}$\ \ \ $4.8\times 10^{-4}$ & $5\times10^6$ &87\ \ \ \ \ \ \ \ 3& $10^6$\\
\hline
\rule{0pt}{12pt}$20\ \ \ \ 1024$ & $1/8$ & $10\ \ 15$ & 470\ \ \ \  413&\sqz$9.5\times 10^{-4}$\ \ \ $8.0\times 10^{-4}$ & $5\times10^6$ &55\ \ \ \ \ \ \ 32 & $10^6$\\
\hline
\rule{0pt}{12pt}$20\ \ \ \ 1024$ & $1/8$ & $8.6\ \ 11.8$ &620\ \ \ \ 480 &\sqz$7.2\times 10^{-4}$\ \ \ $5.6\times 10^{-4}$ & $1\times10^7$ &42\ \ \ \ \ \ \ 20& $10^6$\\
\hline
\hline
\end{tabular}
\caption{Summary of the reduced bases constructed with \pyroq for the {\tt IMRPhenomXPHM} waveform model. The first three sets of bases are in the parameter ranges that the model was compared against the existing NR waveforms, limiting the chirp mass $26\ M\odot\leq M_c \leq 110\ M\odot$, the mass ratio $1\leq q \leq 4$, the magnitudes of the two spins $-0.8\leq \chi_i\leq0.8$ for $i \in [1,2]$, and the full range for the spin angles $(0,0) \leq (\theta_J , \alpha_0) \leq (\pi, 2\pi)$. The other sets all have the same parameter ranges as those of the first three basis sets except the chirp mass range.\\
* For those values the condition $\sigma_{\text{EI}}>10^{-5}$  is used.
}
\label{tab:roq-imrphenomxphm}
\end{table*}

Figure~\ref{fig:xphmempiricalinterpolant} shows an instance of the empirical interpolation for an {\tt IMRPhenomXPHM} waveform and its self-overlap using the last set of the bases in Table~\ref{tab:roq-imrphenomxphm}. The chirp mass is $\mathcal{M}_c=8.8\ M_{\odot}$ and the mass ratio is 1.8. 
Unlike the {\tt IMRPhenomPv2} waveform self-overlaps (see Fig.~\ref{fig:empiricalinterpolant} for an example), there are many features such as those small wiggles in the {\tt IMRPhenomXPHM} waveform self-overlaps as can be seen from Fig.~\ref{fig:xphmempiricalinterpolant}. With other parameters being the same, the larger the mass ratio, the more prominent the features are. For lower chirp mass values, the quadratic basis size becomes more comparable to the size of the corresponding linear basis as shown in Table~\ref{tab:roq-imrphenomxphm}; also refer to the two sets of bases 4s-P and 8s-P in Table~\ref{tab:4secondbasescomparison}.

\subsection{Application of ROQ method to static LISA}
\label{subsec:lisa}
%(There should be a paragraph in the Introduction section about LISA parameter estimation with the ROQ method. \hq{Should we mention the student intern?}) \vflr{That's a very good section to add *if* we have some sort of hard result, i.e. a plot of a ROQ run at LISA's frequencies. If not, we should not mention this, and perhaps add it after we do have something.}
We briefly introduce an application of the ROQ method to a simulated toy detection scenario with the space detector LISA. First, with \pyroq we built the ROQs for days-long ($ M_c \in [10^6,10^7]\ M_{\odot}$) signals of supermassive black hole binaries using the {\tt IMRPhenomPv2} waveform model. %The frequency node separation is $1/1638400-second$ (about once per 19 days).  
\iffalse
\begin{figure}[htp]
    \centering
    \includegraphics[height=7.5cm]{lisa-example-interpolation.png}
    \caption{Empirical interpolants of a random 1638400-second (about 19-day) test waveform (top panel) and its overlap with itself (bottom panel). The chirp mass is $2.8\times 10^6 M_\odot$, and the mass ratio is $1.25$. The light blue thick lines are the original waveform (top) and the original overlap (bottom). Their corresponding interpolants are shown in deep blue dashed lines. The red dots are the empirical nodes.}
    \label{fig:lisaempiricalinterpolant}
\end{figure}

It is shown in Fig.~\ref{fig:lisaempiricalinterpolant} the empirical interpolants for a randomly chosen example waveform in the parameter range in which the ROQs were built.
A randomly chosen waveform in the parameter ranges in which the ROQs were built has the chirp mass $2.8\times 10^6 M_\odot$, up to a factor of $10^5$ than the first detected BBH, and the mass ratio $1.25$.\fi
With \bilby, we created a detector that has an arm length of $2.5\times 10^{-6} km$, which is about the size of LISA, and set it to be located on the surface of the Earth and have the PSDs similar to that of LISA. We call it static LISA, in the sense that we ignore the motion of LISA throughout a signals duration, which can be days, months, and even years. Then we made injections of SMBHB mergers into the static LISA detector and recover the gravitational wave parameters using both the standard method and the ROQ method. The overall speedup was about 5 times using the ROQ method for all the injections and detections of weeks-long signals we simulated. %For the example frequency-domain signal that spans a few mHz, the signal size is actually pretty small with $\sim 10^4$ data points. 
For years-long signal in LISA, greater inference speedup is expected from the ROQ method. We estimate that there will be non-negligible speedup for the LISA parameter estimation with the ROQ method using the time delay interferometery (TDI; see \citep{PhysRevD.65.082003,Vallisneri:2020otf,Tinto:2020fcc}) data collected by the moving LISA.

%%%%%%%%%%%%%%%%%%%%%%%%%%%%%%%%%
\section{Conclusion}
\label{sec:conclusion}
%%%%%%%%%%%%%%%%%%%%%%%%%%%%%%%%%
We have described a novel algorithm, and its code implementation \pyroq, for building reduced bases of gravitational waveforms for fast and accurate gravitational wave inference. 

The code was first validated against the existing and ground-breaking offline ROQ building code \greedycpp with the {\tt IMRPhenomPv2} model, which is a most widely used waveform model in the LVK parameter estimation. We find that \pyroq can build reduced bases of significantly smaller sizes while performing more accurate waveform interpolation compared to \greedycpp, see Table~\ref{tab:4secondbasescomparison}. The computing time is on the scale of a few days. We also showcase how accurate the \pyroq built bases can interpolate test waveforms in Figs.~\ref{fig:empiricalinterpolant} and \ref{fig:waveformerrors}.

Subsequently, point-by-point likelihood comparisons for zero-noise NSBH injections are made between the standard full likelihood and the ROQ likelihood functions with \pyroq built ROQs for {\tt IMRPhenomPv2}. The fractional differences are less than $10^{-3}$ as shown in Fig.~\ref{fig:likelihoodcomparison} for an example injection. Further, recoveries of the simulated zero-noise NSBH injections using \bilby show that gravitational wave inference using \pyroq built ROQs yields visually identical posterior distributions compared with the standard method, see Fig.~\ref{fig:pecomparisonfornsbhinjection}. 

The ROQs produced with \pyroq are also used to infer the first gravitational wave detection GW150914. The parameter estimation results are barely distinguishable compared to the standard method, see Fig.~\ref{fig:gw150914comparisons}. We conclude that \pyroq can build quality waveform bases for waveform models to be more efficiently used in the LVK parameter estimation. 

The reduced bases for {\tt IMRPhenomXPHM}, currently one of the most advanced BBH waveform models that include subdominant harmonic modes and precessing spins, are first built and listed in Table~\ref{tab:roq-imrphenomxphm} as a concrete example of \pyroq's capacity. As the model takes about four times longer in likelihood evaluation than {\tt IMRPhenomPv2}, these bases are expected to substantially reduce the computational costs spent on the parameter estimation of gravitational waves.   

Since the ROQs are built without any noise, they can be used for other ground-based gravitational wave detectors of similar sensitivity range to the LVK detectors. The ROQs can also be used in science scenario studies where simulated gravitational wave signals are injected into the PSDs of the current and the future LVK observations. They are particularly useful when the studies incorporate a large number of simulated events where computation speed becomes an demanding requirement.

It is worth noting that \pyroq can be easily adapted to build ROQ data for any other gravitational waveform models that are not currently included or released in \lalsuite. In addition, we showcase that \pyroq can be adapted for use of space detectors such as LISA, see Sec.~\ref{subsec:lisa}. It can also be modified to work for more generalized time and frequency series from experiments in other research fields that require faster inference.

\begin{acknowledgments}
We wish to acknowledge the useful discussions with Michael P\"urrer,  Rory Smith, Carl-Johan Haster, and Scott Field. We thank Zu-Cheng Chen for careful code review and considerable code improvement. HQ particularly thanks the members in 53 the Parade for brainstorming up the neat name \pyroq in the spring of 2019. We also thank our LIGO Scientific Collaboration Presentation and Publication reviewers for their feedback on this manuscript.
The authors are grateful for the computational resources provided by the LIGO Laboratory and supported by National Science Foundation Grants PHY-0757058 and PHY-0823459. We are grateful for the computational resources provided by the Leonard E. Parker Center for Gravitation, Cosmology and Astrophysics at University of Wisconsin-Milwaukee and supported by NSF-0923409. We thank the Hawk supercomputing system provided by Cardiff University. The project is supported by the STFC grant ST/T000147/1. 
\end{acknowledgments}

%\appendix
%\section{Appendixes}
%To start the appendixes...

\bibliography{pyroq}% Produces the bibliography via BibTeX.

\end{document}